# Terahertz conductivity of Si and of Ge/Si(001) heterostructures with quantum dots

E.S. Zhukova, B.P. Gorshunov, A.S. Prokhorov, I.E. Spektor, Yu.G. Goncharov, L.V. Arapkina,
V.A. Chapnin, V.P. Kalinushkin, G.N. Mikhailova, and V.A. Yuryev
A.M. Prokhorov General Physics Institute, 38 Vavilov street, 119991 Moscow, Russia.

*Abstract*—**With an MBE technique, a Si/Ge heterostructures are prepared containing layers of nanostructured Ge with quantum dots of size of several nanometers. The effective conductivity of the layers is determined by a quasioptical terahertz-subterahertz coherent source BWO spectroscopy. The conductivity is found to be strongly enhanced compared with the conductivity of bulk germanium. Possible microscopic mechanisms responsible for the enhancement will be discussed. Application of BWO spectrometers for obtaining precise quantitative information on of dielectric properties at THz-subTHz frequencies of semiconducting layers and structures is demonstrated by presenting the temperature dependences of dielectric characteristics of a commercial silicon wafer at frequencies 0.3 to 1.2 THz and temperatures 5 K – 300 K.**

Nanoscaled semiconductor-based structures attract presently a lot of interest because of promising practical applications. This stimulates active research in the field of fundamental electronic properties of nano-objects of various dimensionality—quantum dots, nano-wires, two-dimensional structures—as well as of arrays of the elements. Besides, the needs of modern microelectronics require information on the dielectric properties (dielectric permittivity—real and imaginary parts, $\varepsilon'$ and $\varepsilon''$, respectively) of semiconductor materials and structures in the terahertz and subterahertz (THz, subTHz) range (frequencies $\nu \leq 1$ THz). There is no such data in literature. The most precise quantitative information of this type is provided nowadays by quasioptical THz-subTHz spectrometers based on monochromatic and frequency-tunable sources of radiation—backward-wave oscillators (BWOs) [1,2]. In general, these spectrometers allow for direct and, importantly, contactless and polarization-sensitive measurements of the spectra of complex dielectric permittivity $\varepsilon^*(\nu)=\varepsilon'(\nu)+i\varepsilon''(\nu)$ at frequencies from 0.03 up to 1.5 THz, in the temperature interval 2 to 1000 K and in magnetic fields (when needed) up to 8 Tesla (up to 20 Tesla in pulsed fields). In this communication, we present our preliminary results of the study of dynamic conductivity mechanisms of an ensemble of Ge-quantum dots. We have already successfully used this technique to study molecular magnetic nano-clusters [3] and nano-porous silicon structures [4]. We also demonstrate the capabilities of BWO-spectrometers for characterization of bulk semiconducting materials taking the commercial wafers of boron doped silicon with resistivity 12 Ohm*cm (KDB-12).

Using ultra-high vacuum molecular beam epitaxy (UHV MBE) technique, we have prepared a number of heterostructures containing germanium quantum dots and undoped silicon layers on a specially prepared [5] commercial B-doped ($\rho = 12$ Ω·cm) (100)-oriented silicon wafers (KDB-12). Each structure contained a Si buffer layer (thickness 100 nm) deposited on a substrate and a Ge layer whose effective thickness (i.e. the thickness measured by the quartz film thickness monitor, $h_{Ge}$) was varied from 4 Å to 18 Å in different samples. On the top of the Ge layer, another Si layer of thickness of 50 nm was grown and the resulting double-structure was repeated 5 times and finally protected on the top by a cap 50-nm thick Si film. The growth temperature was 560°C for Si layers and 375°C for Ge ones. The samples were studied *in situ* with UHV STM connected to the MBE chamber through a sample transfer line. The STM allowed us to reveal the microstructure of the Ge layers and to determine the the morphology, parameters and concentration of Ge nano-dots. An example of the structure is given in Fig.1 which shows that the Ge layer typically consisted of an array of hut clusters shaped as a wedge or a pyramid. The concentration of quantum dots varied from $2\times10^{11}$ to $6\times10^{11}$ cm$^{-2}$, their height and the base width changed from ~ 0.7 to ~ 1.5 nm and from ~ 5 to ~ 15 nm, respectively, depending on $h_{Ge}$.

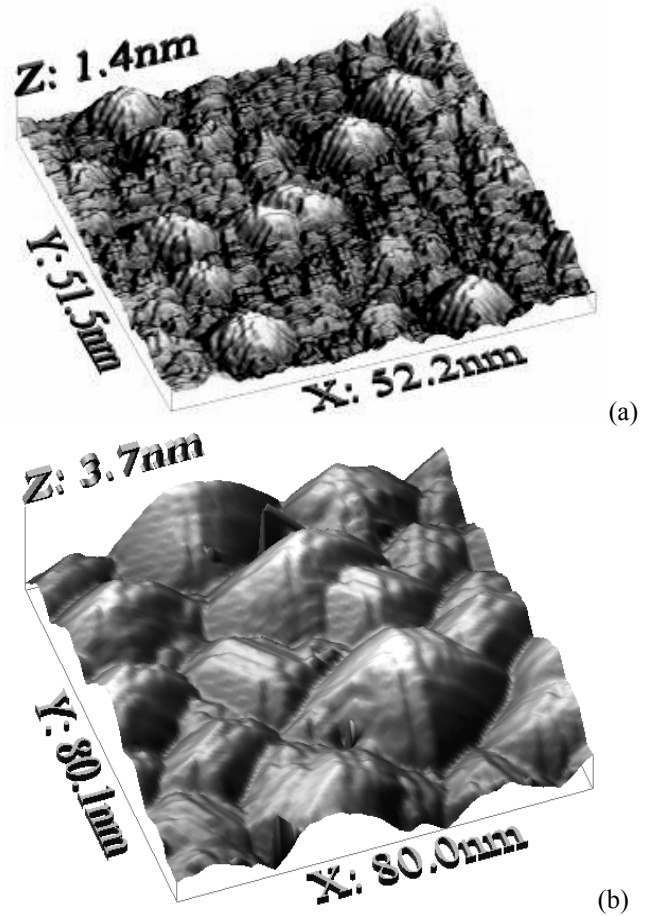

*Fig.1.* STM images of a Si(001) surface with Ge-quantum dots layers: $h_{Ge}$ = 6 Å (a) and 14 Å (b).

The THz-subTHz measurements were performed at room temperature in the range of frequencies 200 GHz to 1 THz. In a quasioptical configuration, the transmission coefficient was measured of a multilayered system "Si substrate + Si/Ge structure". The THz dielectric parameters (real and imaginary parts of the dielectric permittivity) of the substrate were determined beforehand. Comparing transmission coefficients of the samples with that of a bare substrate and using expressions for transmission coefficient of a multilayer system made it possible to determine the effective electrodynamical characteristics of an individual Ge nano-structured layer—conductivity and charge carriers' plasma frequency and scattering rate. Our main preliminary result is that the effective conductivity value, obtained for several samples with different $h_{Ge}$, is unexpectedly high—of an order of 100 Ohm$^{-1}$cm$^{-1}$—when compared with the conductivity of "bulk" Ge (about $10^{-2}$ $\Omega^{-1}$cm$^{-1}$ at room temperature). We discuss possible microscopic mechanisms which could be responsible for the conductivity enhancement.

In Fig.2 we present the results obtained for the KDB-12 silicon substrate. The figure shows the temperature dependences of the parameters of the Drude conductivity model [6] which describes the dynamical response of KDB-12 Si at frequencies from 0.03 to 1.2 THz. Within this model the complex conductivity $\sigma^*$ of a conductor is expressed as:

$$\sigma^*(\nu) = \frac{\sigma_0}{\left(1 - i\frac{\nu}{\gamma}\right)},$$

where $\sigma_0$ is the DC conductivity ($\rho_0 = 1/\sigma_0$ is the DC resistivity) and $\gamma$ is the relaxation frequency; the carriers' plasma frequency is expressed as $\nu_{pl}=(ne^2/m\pi)^{1/2}$ (here $m$ is the effective mass and $n$ is the concentration).

By using a specially designed unit (two-dimensional, 2D, scanner) the spectrometer allows for spatial characterization of semiconductor wafers by performing a two-dimensional probing the surface with a focused beam of the THz radiation. The area covered during the 2D scanning process is up to 55×55 mm$^2$, the step is 30 μm, the scanning speed is 20 mm/s and the spatial resolution is determined by a radiation wavelength. We mention that basing on BWOs, a THz spectroscopy and imaging with a sub-wavelength resolution can be realized [6].

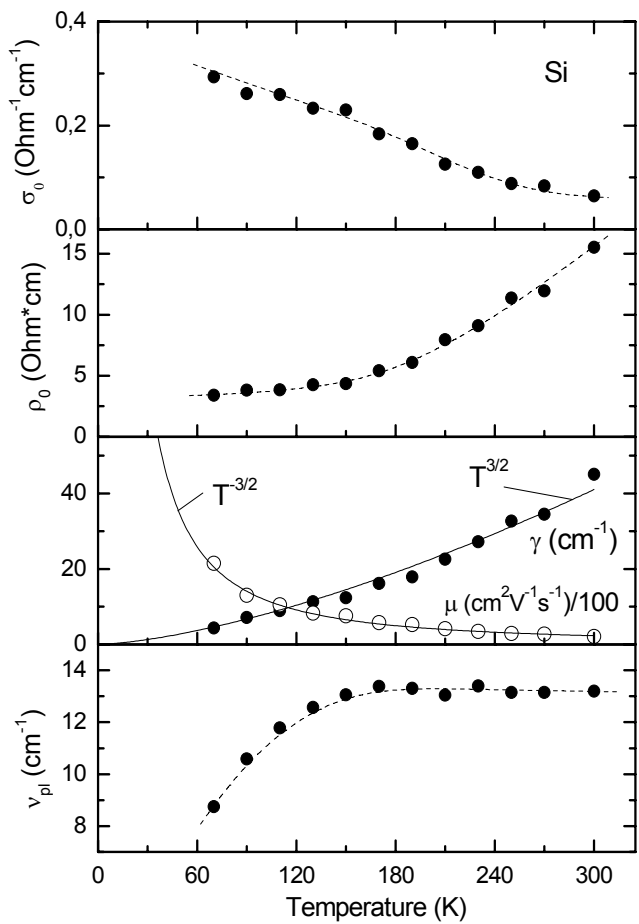

*Fig.2.* Temperature dependences of the parameters of the Drude conductivity model [5] describing the electrodynamic response of the commercial KDB-12 silicon wafer at frequencies 0.3 THz – 1.2 THz. Dashed lines are guides for the eye.